\documentclass[journal]{IEEEtran}
\ifCLASSINFOpdf

\else

\fi
\usepackage{amsmath}
\usepackage{graphicx}
\usepackage{subfig}
\usepackage{diagbox}
\usepackage{caption}
\usepackage{epstopdf}
\usepackage{cite}
\usepackage{amsfonts}
\usepackage{amssymb}
\usepackage{algorithmic,algorithm}
\usepackage{xcolor}
\begin{document}

\title{{C}odeword {P}osition {I}ndex based {S}parse {C}ode {M}ultiple {A}ccess {S}ystem}

\author{Ke Lai, Lei Wen, Jing Lei,  Gaojie Chen \IEEEmembership{Senior Member,~IEEE}, Pei Xiao \IEEEmembership{Senior Member,~IEEE} and Amine Maaref \IEEEmembership{Senior Member,~IEEE}
	\thanks{K. Lai, L. Wen, J. Lei are with Department of Communication Engineering, College of Electronic Science and Engineering, National University of Defence technology. L. Wen is also with the Institute for Communication Systems (ICS), Home of the 5G Innovation Centre (5GIC), University of Surrey, Guildford GU2 7XH, U.K. E-mail: newton1108@126.com }
	\thanks{G. Chen is with the Department of Engineering, University of Leicester, Leicester LE1 7RH, U.K. E-mail: gaojie.chen@leicester.ac.uk.}
	\thanks{P. Xiao is with the Institute for Communication Systems (ICS), Home of the 5G Innovation Centre (5GIC), University of Surrey, Guildford GU2 7XH, U.K. Emails: p.xiao@surrey.ac.uk.}
	\thanks{A. Maaref is with the Huawei Technologies in Ottawa, ON, Canada. Emails: Amine.Maaref@huawei.com.}
	
}
\maketitle

\begin{abstract}
In this letter, a novel variation of sparse code multiple access (SCMA), called codeword position index based SCMA (CPI-SCMA), is proposed. In this scheme, the information is transmitted not only by the codewords in $M$ point SCMA codebook, but also by the indices of the codeword positions in a data block. As such, both the power and transmission efficiency (TE) can be improved, moreover, CPI-SCMA can achieve a better error rate performance compare to conventional SCMA (C-SCMA).
\end{abstract}


\section{Introduction}
\IEEEPARstart {S}{C}{M}{A} \cite{Hosein2013SCMA} is a code domain non-orthogonal multiple access(CD-NOMA)  scheme that is considered to be a promising 5G candidate due to its strong ability to support massive number of users/devices under heavily loaded conditions. In SCMA, the information bits are assigned to multidimensional complex codewords that are not orthogonal to each other using a predefined codebook. The received superimposed signals can be detected via message passing algorithm (MPA).

Since the transmission efficiency (TE) is fixed in C-SCMA systems and the application scenario of C-SCMA such as Internet of Things (IoT) requires flexible data rate under different situations. Moreover, the error rate performance of C-SCMA deteriorate rapidly in fading channels. To make C-SCMA systems more adaptable according to data rate and further improve the error rate performance, we propose to introduce the mechanism of index selection in the codeword position of C-SCMA. Similar idea has been applied to multi-antenna system, i.e., spatial modulation (SM) \cite{Mesleh2008SM} which exploits the spatial domain degree of freedom by conveying information with the indices of the transmit antennas (TAs), and the application of the SM principle to the subcarriers of an orthogonal frequency division multiplexing (OFDM) \cite{Ba2013Orthogonal,Tsonev2011Enhanced,Dang2018adaptive}, Some existing works that combine the SM with CD-NOMA and power-domain NOMA (PD-NOMA) have been proposed in \cite{Liu2017Spatial} and \cite{Zhu2017NOMA}, respectively. However, SM amalgamates with NOMA directly in these works, which indicates that the signals of NOMA are transmitted in a multiple-input multiple-output (MIMO) system with the SM principle. 

In this letter, a novel variation of classical SCMA system, i.e., CPI-SCMA, which combines the index selection strategy and SCMA, is proposed. In CPI-SCMA, the transmitted information bits not only map to the $M$ point SCMA codebook, but also map to a look-up table (LUT) which select the activated positions in a CPI-SCMA codeword. As such, the index of the activated positions in a CPI-SCMA codeword can also carry information. In contrast to the existing works, the proposed CPI-SCMA merge the index selection techniques into SCMA, which to the best of the authors' knowledge, has not been reported in the existing literature. Moreover, CPI-SCMA can achieve a better error rate performance than C-SCMA especially in the region of high signal-to-noise ratio (SNR). Since the transmitted position indices and their corresponding data are both transmitted via superimposed signals, a novel detection algorithm with the aid of MPA is also proposed.

\section{System model}
We consider the uplink transmissions where $J$ single-antenna users transmit to the same base station (BS) with $K$ allocated resources. In C-SCMA, $\log_2M$ bits of the user $j$ directly map to a $K$-dimension codeword vector $\boldsymbol{x_j}$. However, the transmitted $m$ bits for each data block are divided into two parts in CPI-SCMA, the first $m_1$ bits are used to select the active positions from the predefined LUT, and the rest $m_2 = t \cdot \log_2M$ bits map to $t$ SCMA codewords, where $t$ is the number of active positions from $n$ available positions. Therefore, $m_1$ can be represented as:
\begin{equation}
m_1 = \lfloor \log_2(C(n,t)) \rfloor, \label{bits_m1}
\end{equation}
where $C(n,t)$ represents the binomial coefficient and $\lfloor \cdot \rfloor$ denotes the floor function. As such, the index space of each data block can be represented as:
\begin{equation}
\mathcal{I} = \{i_1, i_2, \cdots, i_t\}, \label{index_space}
\end{equation}
where $i_{\beta} \in [1, \cdots, n]$, $\beta \in [1, \cdots, t]$. It should be noted that $\mathcal{I}$ is unitary for each user. The transmitted codeword vectors for user $j$ that is generated via SCMA codebook $\mathcal{S}$ can be written as:
\begin{equation}
\boldsymbol X_j =  [\boldsymbol{x_{j,1}}, \boldsymbol{x_{j,2}}, \cdots \boldsymbol{x_{j,t}}]. \label{signal_space}
\end{equation}
Each codeword $\boldsymbol{x_{j,\beta}} \in \mathcal{S}$, where $\beta \in [1, \cdots, t]$, corresponds to the position index $i_{\beta}$ in $\mathcal{I}$, and other inactive positions are set to $\boldsymbol{0}$, which $\boldsymbol{0}$ is a $K$-dimension zero vector. Consequently, the transmitted signal vector $\boldsymbol{c_j}$ for each data block generates from \eqref{index_space} and \eqref{signal_space}.

After $\boldsymbol{c_j}$ is constructed, the codeword of each user serves as the input to the OFDM modulator. Subsequently, the signals of each user are conveyed to Rayleigh fading channel.
After this point, the CPI-SCMA codeword $\boldsymbol{c_j}$ superpose at the receiver, therefore, the received signal at BS can be written as:
\begin{equation}
\boldsymbol y = \sum_{j=1}^J diag(\boldsymbol h_j)\boldsymbol c_j + \boldsymbol z, \label{superpose_sig}
\end{equation}
where $diag(\boldsymbol h_j)$ is the equalized channel matrix, $\boldsymbol z$ is the noise vector that consists of complex Gaussian random variables with distribution $\mathcal{CN}(0,\sigma^2)$, and $\boldsymbol y = [ y_1, \cdots,  y_{n \times K}]$, in which $y_k$ is the $k$th received chip. Since the index selection strategy is applied to the SCMA, the total number of received chips in a data block equals to $n \times K$. It should be noted that the power of $\boldsymbol{c_j}$ has to be normalized to $N$ after OFDM modulation.
For simplicity, we mainly discuss the CPI-SCMA system with $n = 4, t = 2$ in this letter, an example for the codeword position index selector is shown in Table. \ref{LUT}. 

\begin{table}[t!]\scriptsize
	\renewcommand{\arraystretch}{1.3}
	\caption{A look-up table example for $n = 4, t = 2$}
	\label{LUT}
	\centering
	\begin{tabular}{|c||c||c|}
		\hline
		bits &  indices & data block  \\\hline
		
		$[0\quad0]$ & $\{1 \quad3\}$ & [$\boldsymbol{x_{j,1}}$ $\boldsymbol{0}$ $\boldsymbol{x_{j,2}}$ $\boldsymbol{0}$] \\\hline
		
		$[0\quad 1]$ & $\{2 \quad4\}$ & [$\boldsymbol{0}$ $\boldsymbol{x_{j,1}}$  $\boldsymbol{0}$ $\boldsymbol{x_{j,2}}$]\\\hline
		
		$[1 \quad0]$ & $\{2\quad 3\}$ & [$\boldsymbol{0}$ $\boldsymbol{x_{j,1}}$ $\boldsymbol{x_{j,2}}$ $\boldsymbol{0}$] \\\hline
		
		$[1 \quad1]$ & $\{1\quad 4\}$ & [$\boldsymbol{x_{j,1}}$ $\boldsymbol{0}$  $\boldsymbol{0}$ $\boldsymbol{x_{j,2}}$] \\\hline
	\end{tabular}
\end{table}

\section{Detection algorithm}
In this section, a novel message passing aided detection algorithm (MPAD) is introduced to detect the active position indices and the corresponding data. 

\subsection{Error pattern}

To recover the information bits from the received chips in CPI-SCMA, we first analyze the possible error patterns in CPI-SCMA. Unlike C-SCMA, for which information bits can be simply detected via MPA, the effect of both indices and their corresponding information should be taken into consideration in CPI-SCMA. From the analysis of error patterns, the correct codeword can be inferred, and thus significantly reduce the search space of the possible codewords. Taking the CPI-SCMA with parameters $n=4$ and $t=2$ as an example, the error patterns of each user can be classified into the following five cases. For simplicity, we define a candidate set $\psi^{(\gamma)}$ for each case which contains all possible codewords in case $\gamma$. 

\emph{Case 0:} There is no position being detected as $\boldsymbol{0}$, which can be written as:
$ [\boldsymbol{x_{j,1}}  \boldsymbol{x_{j,2}}  \boldsymbol{x_{j,3}}  \boldsymbol{x_{j,4}}]$.
Therefore, all possible candidate codewords can be generated according to the LUT, which indicates that the selected indices are set to active, otherwise inactive. It can be inferred that four elements are included in the candidate set $\psi^{(0)}$.

\emph{Case 1:} Only one position is detected as $\boldsymbol{0}$.
Thus, an extra position in the detected data block should be $\boldsymbol{0}$. According to the LUT, only two possible indices can be selected from $\mathcal{I}$ once one position index is decided. Hence, there are two elements in the candidate set $\psi^{(1)}$ of Case 1.

\emph{Case 2:} There are also two positions which are detected as $\boldsymbol{0}$, however, it is obviously that $C(n,t)-2^{m_1}$ position indices are not included in the LUT that is shown in Table. \ref{LUT}, i.e., $\{1 \quad 2\}$ and $\{3\quad4\}$.
In this case, one of the $\boldsymbol{0}$ positions should be replaced by $\boldsymbol{x_{j,t}}$, furthermore, each position corresponds to two indices; hence, $4\cdot M$ elements should be considered in $\psi^{(2)}$.

\emph{Case 3:} Three positions are detected as $\boldsymbol{0}$. Therefore, one position of $\boldsymbol{0}$ should be replaced by $\boldsymbol{x_{j,t}}$. Similar to Case 1, two possible positions correspond to the detected active position. Moreover, as the $\boldsymbol{x_{j,t}}$ is unknown to the receiver; hence, all possible $M$ symbols should be considered in this case, which leads to a candidate set $\psi^{(3)}$ includes $2 \cdot M$ elements.

\emph{Case 4:} All positions are detected as $\boldsymbol{0}$, which indicates no information can be obtained. Consequently, $2^{m_1} \cdot M^2$ elements construct the set $\psi^{(4)}$.

Note that the error pattern varies with the key parameters in CPI-SCMA, such as $M,n$ and $t$, which indicates that diverse  $\psi^{(\gamma)}$ can be constructed for different CPI-SCMA system. It can be inferred that the total number of error patterns for a given CPI-SCMA system with arbitrary $n$ and $t$ is $I_{C(n,t)-n>0} + n$, where $I_{\mathcal{C}}$ is an indicator function equals to 1 only if condition $\mathcal{C}$ is true. It should be noted that the appearance of error patterns in CPI-SCMA relates to many factors, and thus its analysis from theoretical perspective is necessary for further study.

\subsection{Message passing aided detection algorithm}
As both indices of active codeword positions and the corresponding information should be detected in CPI-SCMA, simple MPA no longer works in this case. To render the system affordable for practical implementation, a novel MPAD is proposed for CPI-SCMA.

 As discussed above, there are $n \times K$ symbols at the receiver, note that each chip is a superimposed signal consists of symbols in $\mathcal{S}$ and zeros. Consequently, the symbols of each user can be detected with the message passing algorithm (MPA) in C-SCMA. However, since inactive positions in $\boldsymbol c_j$ are filled with zeros, the function node (FN) update can be written as:
%\begin{equation}
%\begin{array}{l}
%U_{k \to j}(\tilde{\boldsymbol{x_j}}) = \sum\limits_{\boldsymbol{x} \in \chi} \dfrac{1}{\pi N_0} \exp[-\dfrac{1}{N_0} \vert y_k - h_{j,k}x_{j,k} \\ \quad \quad \quad \quad \quad \quad  \quad \quad 
%- \sum\limits_{i \in \xi_k\backslash j} h_{i,k}x_{i,k}\vert^2] \prod\limits_{i \in \xi_k\backslash j}V_{i \to k}(\tilde{\boldsymbol{x_i}}),\label{FN_updat}
%\end{array}
%\end{equation}
\begin{equation}
\begin{array}{l}
U_{k \to j}(\tilde{\boldsymbol{x_j}}) = \sum\limits_{\boldsymbol{x} \in \chi} \dfrac{1}{\pi N_0} \exp[-\dfrac{1}{N_0} \vert y_k - h_{j,k}x_{j,k}  \\ \quad \quad \quad \quad \quad \quad  \quad \quad 
- \sum\limits_{i \in \xi_k\backslash j} h_{i,k}x_{i,k}\vert^2] \prod\limits_{i \in \xi_k\backslash j}V_{i \to k}(\tilde{\boldsymbol{x_i}}),\label{FN_updat}
\end{array}
\end{equation}
where $\tilde{\boldsymbol{x_j}}$ is the detected codeword for user $j$, $y_k$ is the $k$th component of $\boldsymbol{y}$, $\xi_k$ is a node set that contains all the user nodes (UNs) that are connected to the $k$th FN, and $\chi$ is a combination set consists of $M+1$ elements that is defined by:
\begin{equation}
\begin{array}{l}
\chi(M+1) = \{\tilde{\boldsymbol{x_k}}=[0,x_1^{k_1},\cdots,x_{d_f}^{k_{d_f}}]:\\
\forall \boldsymbol k=[k_1,\cdots,k_{d_f}]\in\{0,1,\cdots,M\}^{d_f}\}, \label{COMB_SCMA}
\end{array}
\end{equation}
where $d_f$ is the degree of FN. It should be noted that an extra symbol 0 is added in the MPA, which follows from the fact that the inactive positions of $\boldsymbol{c_j}$ are filled with $\boldsymbol{0}$; hence, the indices of inactive positions can be detected.
After the MPA processing, the index of the active positions in the CPI-SCMA codeword can be obtained, however, as the error symbols exist, the indices and symbols solely detected by MPA are not sufficient to recover the data block, especially in the low SNR region.

As can be observed from \eqref{superpose_sig}, once $\boldsymbol{c_j}$ is detected correctly in MPA, it can be canceled from the superimposed signals to eliminate the interferences. In this letter, the detected codewords with formation in Table. \ref{LUT} are regarded as "reliable" codewords. The rest of the superimposed signals can be represented as:
\begin{equation}
\boldsymbol{r} = \boldsymbol{y} - \sum_{j \in \boldsymbol{u}, k = 1}^{n\times K}c_{j,k}h_{j,k}\label{rest_sig}
\end{equation}
where $\boldsymbol{u}$ is a set that contains users that are detected as reliable codewords with modified MPA shown in \eqref{FN_updat} and \eqref{COMB_SCMA}. After the  cancellation of reliable codewords, $\boldsymbol{r}$ consists of the elements in candidate set $\psi^{(\gamma)}$, therefore, we can detect the rest users' data with minimum Euclidean rule:
\begin{equation}
\hat{\boldsymbol{c_j}} = \text{arg} \min_{\boldsymbol{c_j}\in \boldsymbol{\Psi}} \Vert \boldsymbol{r} - \sum_{j \notin \boldsymbol{u}}\boldsymbol{c_j}\boldsymbol{h_j} \Vert, \label{ML}
\end{equation}
where $\hat{\boldsymbol{c_j}}$ is the tentative estimate of the codeword, and $\Psi$ is the Cartesian product of candidate set $\psi^{(\gamma)}$, which can be written as:
\begin{equation}
\Psi = \mathcal{F}(\psi^{(1)}) \times \cdots \times \mathcal{F}(\psi^{(\gamma)})
\end{equation}
Note that $\mathcal{F}$ is an indicator operator that can be defined as:
\begin{equation}
\mathcal{F}(\psi^{(\gamma)})= \left\{ \begin{array}{ll}
\varnothing & \textrm{if error pattern $\gamma$ does not exist}\\
\psi^{(\gamma)} & \textrm{if error pattern $\gamma$ exist}\\
\end{array} \right.\label{ind} 
\end{equation} 
From \eqref{ML}, it can be observed that this step is essentially a partial ML (PML) detector, which should be attributed to the analysis of error patterns. As such, the cardinality of $\psi^{(\gamma)}$ is far lower than the search space in ML. Therefore, the proposed MPAD is a four steps detector with the aid of MPA.
To sum up, the MPAD can be summarized as Algorithm 1.
\begin{algorithm}
	\caption{MPAD detection for CPI-SCMA}

	\begin{algorithmic}[1]  
		\STATE \textbf{Inputs:} $y_k,h_{j,k},M,n,t,\psi^{(\gamma)},LUT$
			 \STATE  \textbf{Modified MPA:}\\
		\FOR {$\text{iter}=1,\cdots,\text{iter\_num}$}	
		\STATE \textbf{Initialization:} \\
		$V_{j \to k}(\tilde{\boldsymbol{x_j}})=0,\tilde{\boldsymbol{x_j}}\in \chi(q)$

		\STATE \textbf{FN update:} \\	
		Using \eqref{FN_updat}, the definition of $\tilde{\boldsymbol{x_j}}$ is the same as \eqref{COMB_SCMA}.
		
		\STATE \textbf{UN updat:} \\	
		$V_{j \to k}(\tilde{\boldsymbol{x_j}})=\prod\limits_{l\in \xi_j \backslash k}U_{l \to j}(\tilde{\boldsymbol{x_j}})$.
		
		\STATE  \textbf{Decision:}\\
		$V_{j}(\tilde{\boldsymbol{x_j}}) = \prod\limits_{k\in \xi_j}U_{k \to j}(\tilde{\boldsymbol{x_j}})$
				
		\ENDFOR
		
	   \STATE  \textbf{Codeword cancellation (CC):}\\
	   1. Selecting the decided codewords with the same formation in LUT and their corresponding users $\boldsymbol{u}$.\\
	   2. Constructing $\psi^{(\gamma)}$ according to the error patterns.\\
	   3. Using \eqref{rest_sig} to process CC, and obtain $\boldsymbol{r}$.
	   
	   \STATE  \textbf{Partial ML detection (PML):}\\
	   Processing \eqref{ML} in the user set $\complement_{\boldsymbol{J}}\boldsymbol{u}$, where $\complement_{\boldsymbol{J}}\boldsymbol{u}$ denotes the complementary set of $\boldsymbol{J}$.\\
	   \STATE  \textbf{Final decision:}\\
	   Combining the results in Modified MPA and PML detection.
	   
	\end{algorithmic}
\end{algorithm}
\section{Numerical results and analysis}
In this section, the simulation results are presented to demonstrate the effectiveness of the proposed scheme. The simulation parameters are set to: $J=6, K=4, M=4$, the index selector is the same as Table. \ref{LUT}.  The CPI-SCMA use the same codebook as C-SCMA, which is designed according to \cite{Taherzadeh2014SCMA}.
 We define the TE ($r_t$) as the ability to carry bits of each chip (physical resource), and thus the $r_t^{c} = J\cdot \log_2M/K = 3$ bits/chip for C-SCMA. As for CPI-SCMA, the TE can be represented as:
\begin{equation}
r_t^{cpi}= J(m_1 + m_2)/(nK).  
\end{equation}
Therefore, the condition to achieve a higher TE is $r_t^{cpi}/r_t^{c} \geq 1$.

 \begin{figure}[h]
	\centering

	\renewcommand{\captionfont}{\small } \renewcommand{\captionlabelfont}{\small} \centering \includegraphics[height=2.0in, width=2.6in]{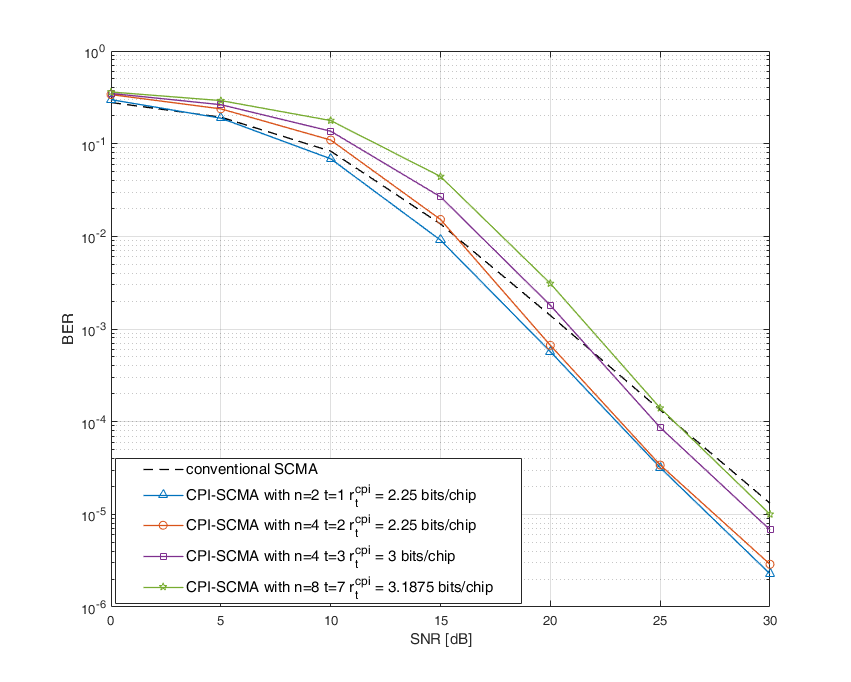}\caption{Performance comparisons of CPI-SCMA and C-SCMA.}

	\label{fig_CIS_SCMA}

\end{figure}
The error rate performances of CPI-SCMA and C-SCMA are evaluated in Fig. \ref{fig_CIS_SCMA}. It can be observed from the figure that the bit error rate (BER) of CPI-SCMA can outperform C-SCMA in several cases. When the TE of CPI-SCMA is lower than C-SCMA, the BER performance of CPI-SCMA can be significant improved, which brings about approximate to 4 dB gain at BER = $10^{-4.9}$.
We also simulate the CPI-SCMA with the same TE as C-SCMA, for which $n=4, t=3$. As can be seen from the figure, the gain of CPI-SCMA with $r_t^{cpi} = 3$ bits/chip decreases to 2 dB at BER = $10^{-4.9}$ compare to CPI-SCMA with $r_t^{cpi} = 2.25$ bits/chip, and the performance is worse than C-SCMA at low SNRs. However, the BER curve cross the C-SCMA curves in the region of high SNRs. The performance with higher TE is depicted in the figure as well. As shown in the figure, the crossover point of CPI-SCMA and C-SCMA moves down, which indicates that a better error rate performance compare to C-SCMA can be achieved by CPI-SCMA with $n=8, t=7$ at SNR$>$25 dB. Furthermore, when TE is the same, such as $n=2, t=1$ and $n=4, t=2$, CPI-SCMA with smaller $n$ and $t$ performs better in the region of low SNRs.

\begin{figure}[h]
	\centering
	
	\renewcommand{\captionfont}{\small } \renewcommand{\captionlabelfont}{\small} \centering \includegraphics[height=2.0in, width=3in]{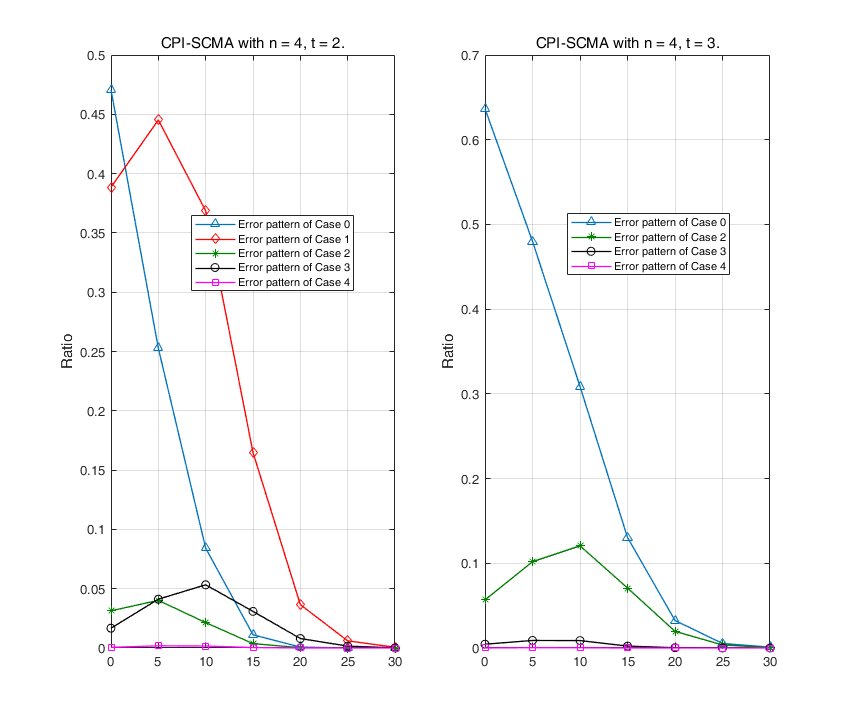}\caption{Ratio of different error patterns in CPI-SCMA.}
	
	\label{fig_error_pattern_ratio}
	
\end{figure}

The ratio of different types of error patterns in different CPI-SCMA systems, which is defined as $\delta_{\gamma} = I_{\mathcal{F}(\psi^{(\gamma)})}/\mathcal{B}$, are depicted in Fig. \ref{fig_error_pattern_ratio}, where $I_{\mathcal{C}}$ is an indicator function that has been defined in previous, $\mathcal{B}$ is the number of total simulated blocks. For the case when $n = 4, t = 2$, it is shown that the Case 0 and Case 1 analyzed in Sec. II have the largest probability to happen for CPI-SCMA with $n = 4, t= 2$, however, the ratio of Case 0 decrease more rapid than Case 1. We have shown that the cardinality of $\vert \psi^{(0)}\vert$ and  $\vert \psi^{(1)}\vert$ equals to 4 and 2, respectively, which are the lowest. Furthermore, the ratio of Case 4 is close to 0, and the ratio of Case 2 and Case 3 do not exceed 0.05 at any SNRs.  As for CPI-SCMA with $n = 4, t= 3$, there are merely four different types of error patterns since all the combinations can be used to select indices in the LUT, and the $\psi^{(0)}$ that possess the highest ratio among all error patterns is also with the lowest cardinality, which equals to 4. The ratio of $\vert \psi^{(2)}\vert = 8$ is very low at any SNRs compare with the rest patterns that are with high cardinality. The poor performance at low SNRs can be explained by the fact that the total amount of reliable codewords are much smaller than the erroneous codewords and the probability to correctly decode a codeword is low, which may lead to a deviation and error propagation in CC; hence, the PML is unable to recover the information of each user with high reliability. 
\begin{table}[t!]\scriptsize
	\renewcommand{\arraystretch}{1.3}
	\caption{Average extra complexity by using PML in CPI-SCMA}
	\label{complexity}
	\centering
	\begin{tabular}{|c||c||c||c|}
		\hline
     	\diagbox{system}{$E_b/N_0$} & 0 dB & 15 dB  & 30 dB\\\hline		
		$n=4,t=2$ & 3.33 & 1.95 & $2.2\times 10^{-3}$\\\hline		
		$n=4,t=3$ & 3.24 & 2.69 & $7.8\times 10^{-3}$\\\hline		
	\end{tabular}
\end{table}

A good approximation for the $j$th user's average block error rate (ABLER) can be given by:
\begin{equation}
P_j(e) \leq \dfrac{1}{(M^t\cdot n)^J} \sum_{\mathbf{C}} \left( \sum_{\mathbf{\hat{C}},\mathbf{c^j \neq \hat{c}^j}} P(\mathbf{C} \to \mathbf{\hat{C}})\right),
\end{equation}
where $\mathbf{C}$ and $\mathbf{\hat{C}}$ are the transmitted and the erroneously detected codeword set for all users, respectively. A good approximation of $Q$-function is $Q(x) \approx 1/12e^{-x^2/2}+1/4e^{-2x^2/3}$ \cite{Chiani2003New}, therefore, the unconditional pairwise error probability (UPEP), i.e., $P(\mathbf{C} \to \mathbf{\hat{C}})$ can be further written as:
\begin{equation}
\begin{array}{rcl}

P(\mathbf{C} \to \mathbf{\hat{C}}) &=& \mathbb{E}_\mathbf{h}\{P(\mathbf{C} \to \mathbf{\hat{C}}|\mathbf{h})\}\\
&\approx& \mathbb{E}_\mathbf{h}\{\dfrac{1}{12}\exp(-\dfrac{\sum_{s=1}^{n \times K}\lambda_{s}^2\vert h_s\vert^2}{4N_0} )\\
&+&\dfrac{1}{4}\exp(-\dfrac{\sum_{s=1}^{n \times K}\lambda_{s}^2\vert h_s\vert^2}{3N_0})\}\\
 &=&\prod_{s=1}^{n \times K} \dfrac{N_0(48N_0-13\lambda_s^2)}{(4N_0-\lambda_s^2)(6N_0-\lambda_s^2)},
\end{array}
\end{equation}
where $\lambda_s^2 = \sum_{j=1}^J\vert c^j[s] -  \hat{c}^j[s]\vert^2$, denoting the differences between transmitted and erroneously detected codewords.

The complexity of modified MPA in MPAD is in the order of $\mathcal{O}(M+1)^{d_f}$, which is slightly higher than the original MPA.  As can be seen from Algorithm. 1, the average extra computational complexity of CPI-SCMA mainly lies on PML detection. Based on Fig. 2, extra searching space by using PML is included in Table. \ref{complexity}. As shown in Table. \ref{complexity}, the extra searching space of the C-SCMA is defined as 0, and the data in the table is the relative complexity calculated by $\sum_{\gamma}\vert \psi^{(\gamma)}\vert \delta_{\gamma}$ compare to the the C-SCMA. It is clear that the complexity of MPAD decrease as SNR increasing, and the extra complexity of MPAD gradually approximates to 0. Therefore, it can be inferred that the complexity of MPAD is much lower than the ML detector and approximates to that of the original MPA for the C-SCMA, especially at high SNRs. 

\section{Conclusion}
In this letter, a novel variation of C-SCMA that introduces the codeword position index selection strategy, is proposed. To detect the information of CPI-SCMA, a novel MPAD with modified MPA and CC is presented. Our future work aims to resolve the following
	open issues: (1) simplified algorithm for CPI-SCMA; (2) analytical results of error patterns; (3) optimization of CPI-SCMA codebook.

\ifCLASSOPTIONcaptionsoff
\newpage
\fi
\bibliographystyle{IEEEtran}
\bibliography{CPI-SCMA}

% Generated by IEEEtran.bst, version: 1.13 (2008/09/30)
\begin{thebibliography}{1}
\providecommand{\url}[1]{#1}
\csname url@samestyle\endcsname
\providecommand{\newblock}{\relax}
\providecommand{\bibinfo}[2]{#2}
\providecommand{\BIBentrySTDinterwordspacing}{\spaceskip=0pt\relax}
\providecommand{\BIBentryALTinterwordstretchfactor}{4}
\providecommand{\BIBentryALTinterwordspacing}{\spaceskip=\fontdimen2\font plus
\BIBentryALTinterwordstretchfactor\fontdimen3\font minus
  \fontdimen4\font\relax}
\providecommand{\BIBforeignlanguage}[2]{{%
\expandafter\ifx\csname l@#1\endcsname\relax
\typeout{** WARNING: IEEEtran.bst: No hyphenation pattern has been}%
\typeout{** loaded for the language `#1'. Using the pattern for}%
\typeout{** the default language instead.}%
\else
\language=\csname l@#1\endcsname
\fi
#2}}
\providecommand{\BIBdecl}{\relax}
\BIBdecl

\bibitem{Hosein2013SCMA}
H.~Nikopour and H.~Baligh, ``Sparse code multiple access,'' in \emph{in Proc.
  IEEE 24th Int. Symp. Pers. Indoor Mobile Radio Commun. (PIMRC)}, Sep. 2013,
  pp. 332--336.

\bibitem{Mesleh2008SM}
R.~Y. Mesleh, H.~Haas, S.~Sinanovic, W.~A. Chang, and S.~Yun, ``Spatial
  modulation,'' \emph{IEEE Transactions on Vehicular Technology}, vol.~57,
  no.~4, pp. 2228--2241, Jul. 2014.

\bibitem{Ba2013Orthogonal}
E.~Başar, Ümit Aygölü, E.~Panayırcı, and H.~V. Poor, ``Orthogonal
  frequency division multiplexing with index modulation,'' \emph{IEEE
  Transactions on Signal Processing}, vol.~61, no.~22, pp. 5536--5549, Nov.
  2013.

\bibitem{Tsonev2011Enhanced}
D.~Tsonev, S.~Sinanovic, and H.~Haas, ``Enhanced subcarrier index modulation
  ({S}{I}{M}) {O}{F}{D}{M},'' in \emph{IEEE GLOBECOM Workshops}, Dec. 2011, pp.
  728--732.

\bibitem{Dang2018adaptive}
S.~Dang, J.~P. Coon, and G.~Chen, ``Adaptive {O}{F}{D}{M} with index modulation
  for two-hop relay-assisted networks,'' \emph{IEEE Transactions on Wireless
  Communications}, vol.~17, no.~3, pp. 1923--1936, Mar 2018.

\bibitem{Liu2017Spatial}
Y.~Liu, L.~L. Yang, and L.~Hanzo, ``Spatial modulation aided sparse
  code-division multiple access,'' \emph{IEEE Transactions on Wireless
  Communications}, vol.~17, no.~3, pp. 1474--1487, Mar. 2018.

\bibitem{Zhu2017NOMA}
X.~Zhu, Z.~Wang, and J.~Cao, ``{N}{O}{M}{A}-based spatial modulation,''
  \emph{IEEE Access}, vol.~5, no.~99, pp. 3790--3800, Apr 2017.

\bibitem{Taherzadeh2014SCMA}
M.~Taherzadeh, H.~Nikopour, A.~Bayesteh, and H.~Baligh, ``{S}{C}{M}{A} codebook
  design,'' in \emph{in Proc. IEEE Veh. Technol. Conf. (VTC-Fall)}, Sep. 2014,
  pp. 1--5.

\bibitem{Chiani2003New}
C.~M, D.~D, and S.~M, K, ``New exponential bounds and approximations for the
  computation of error probability in fading channels,'' \emph{IEEE
  Transactions on Wireless Communications}, vol.~2, no.~4, pp. 840--845, Jul
  2003.

\end{thebibliography}

\end{document}